\documentclass[twocolumn, aps, prd, 
showpacs, floatfix, nofootinbib]{revtex4}

\usepackage[ colorlinks=true, pdfstartview=FitV, linkcolor=red, citecolor=blue, urlcolor=blue]{hyperref}

\usepackage{multirow}
\usepackage{epsfig}
\usepackage{amsmath}

\usepackage{latexsym}
\usepackage{feynmp}
\usepackage[normalem]{ulem}  
\usepackage[dvips]{color} 

\renewcommand\sout{\bgroup \color{red} \ULdepth=-.5ex \ULset}

\newcommand{\vect}[1]{\mathbf{#1}}

\newcommand{\vp}{\vect{p}}
\newcommand{\vk}{\vect{k}}
\newcommand{\vx}{\vect{x}}

\newcommand{\vv}{\vect{v}}

\newcommand{\vzero}{\vect{0}}
\newcommand{\vj}{\vect{j}}
\newcommand{\vA}{\vect{A}}
\newcommand{\vE}{\vect{E}}
\newcommand{\vB}{\vect{B}}

\newcommand{\comment}[1]{}

\newcommand{\Nc}{N_c}
\newcommand{\Nf}{N_f}
\newcommand{\Cf}{C_F}
\newcommand{\nf}{n_F}
\newcommand{\nb}{n_B}

\begin{document}

\title{Exact vector channel sum rules at finite temperature \\
and their applications to lattice QCD 
data analysis}

\author{Philipp Gubler}
\email[]{pgubler@riken.jp}
\affiliation{ECT*, Villa Tambosi, I-38123 Villazzano (Trento), Italy}

\author{Daisuke Satow}
\email[]{dsato@th.physik.uni-frankfurt.de}
\affiliation{ECT*, Villa Tambosi, I-38123 Villazzano (Trento), Italy}

\date{\today}

\begin{abstract}
We derive three exact sum rules for the spectral function of the electromagnetic current with zero spatial momentum at finite temperature. 
Two of them are derived in this paper for the first time. 
We explicitly check that these sum rules are satisfied in the weak coupling regime and examine which sum rule is sensitive to the transport 
peak in the spectral function at low energy or the continuum at high energy. 
Possible applications of the three sum rules to lattice computations of the spectral function and transport coefficients are also discussed: 
We propose an ansatz for the spectral function that can be applied to all three sum rules and fit it to available lattice data of the Euclidean vector correlator above the critical temperature. 
As a result, we obtain estimates for both the electrical conductivity $\sigma$ and the second order transport coefficient $\tau_J$. 
\end{abstract}


\maketitle

\section{Introduction
 and Summary}

Among the properties of hadronic matter at finite temperature, whose dynamics is described by quantum chromodynamics (QCD), 
the spectral function of the electromagnetic current plays an important role since it contains the full information on the dilepton/photon production 
rate~\cite{McLerran:1984ay}, the electrical conductivity, and the modification of the spectral properties of vector mesons at finite temperature. 
All these quantities have been intensively studied in the context of heavy ion collisions. 
The spectral function has therefore naturally been investigated within many approaches, such as perturbative QCD~\cite{Baier:1988xv}, 
the AdS/CFT correspondence~\cite{CaronHuot:2006te}, model calculations~\cite{Gale:2014dfa}, low-energy effective theory based on 
hadronic degrees of freedom~\cite{Chanfray,Klingl}, 
sum rules~\cite{Gubler:2015yna, Huang:1994fs, Kapusta:1993hq, Zschocke:2002mn}, 
and lattice QCD~\cite{Aarts:2007wj, Ding:2010ga, Karsch:2001uw, Bernecker:2011gh, Brandt:2012jc, Brandt:2014qqa, Gupta:2003zh, Burnier:2012ts, Brandt}, 
which have led to a large number of diverse results. 
Under such circumstances, it is useful to have exact constraints on the spectral function that all approaches should satisfy. 
Especially in lattice QCD, which can be directly applicable only for static quantities, it would be useful to have such constraints since the spectral function is a dynamical 
quantity and thus can not be computed directly. 
One goal of the present paper is to provide such constraints in the form of sum rules, and discuss their applications to lattice QCD analysis. 

In the first part of the manuscript, we derive the three sum rules of Eqs.~(\ref{eq:sumrule-1}), (\ref{eq:sumrule-2}), and (\ref{eq:sumrule-3}), 
of which the second and third one are written down here for the first time. 
The third one is valid in the large $\Nc$ limit while the other two are exact for general $\Nc$.
For this purpose, we make use of a method developed for the energy-momentum tensor 
channel in an earlier work by Romatschke and Son \cite{Romatschke:2009ng}. 
We emphasize that these sum rules are exact, and valid both in hadron and quark-gluon plasma phases, as long as hydrodynamics is reliable there. 
We furthermore check that the sum rules are satisfied at weak coupling by explicit perturbative calculations, and examine their sensitivity to the peak in the 
spectral function which is caused by the transport process of the quarks (transport peak) and the continuum generated by free quark pair creation processes. 
Next, we discuss potential applications of the sum rules to lattice QCD studies of the spectral function. 
These include the possibility of providing constraints to the spectral function ansatz used to fit the Euclidean vector correlator lattice data, improvements for this ansatz, and the extraction of the second order transport coefficient $\tau_J$ from the spectral function obtained from a fit to lattice data. 

\begin{figure}[t] 
\begin{center}
\includegraphics[width=0.25\textwidth]{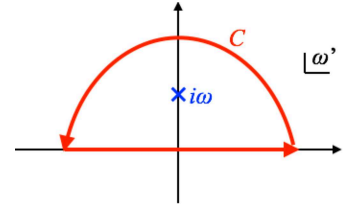} 
\vspace{-13mm}
\caption{The contour $C$, used in the integral of Eq.~(\ref{eq:residue-theorem}).} 
\label{fig:contour} 
\end{center} 
\end{figure} 

\section{Sum rules}
\subsection{Sum Rule 1}

The quantity we are interested in is the retarded Green function of the electromagnetic (EM) current: $G^{R \mu\nu}(\omega,\vp)\equiv i \int dt \int d^3\vx e^{i\omega t-i\vp\cdot \vx}
\theta(t) \langle [j^\mu(t,\vx),j^\nu(0,\vzero)] \rangle$, where $j^\mu \equiv e\sum_f q_f \overline{\psi}_f \gamma^\mu \psi_f $ is the EM current, and the average is taken over the 
thermal ensemble. Here $e$ is the electromagnetic coupling constant, $q_f$ the charge in each quark flavor, and $\psi_f$ the quark field with flavor $f$, respectively.
At $|\vp|=0$, there is only one independent component in the spatial components of this tensor, $G^{R }(\omega)\equiv G^R_{ii}(\omega,\vzero)/3$, due to isotropy.
In this paper, we limit ourselves to this case for simplicity.

First, to introduce the method developed in Ref.~\cite{Romatschke:2009ng}, 
we rederive the sum rule of Eq.~(\ref{eq:sumrule-1}), which has already been obtained in Ref.~\cite{Bernecker:2011gh} from the current conservation law. 
The retarded Green function is known to be analytic in the upper half of the complex $\omega$ plane.
This property enables us to derive various sum rules. Because of the residue theorem, we have
\begin{align}
\label{eq:residue-theorem}
\delta G^{R}(i\omega)-\delta G^{R}_\infty
&= \frac{1}{2\pi i}\oint_C d\omega' \frac{\delta G^{R}(\omega')-\delta G^{R}_\infty}{\omega'-i\omega},
\end{align}
for which the contour $C$ is shown in Fig.~\ref{fig:contour}. 
Here $\delta$ stands for the subtraction of the $T=0$ value of $G^{R}(\omega)$, $\delta G^{R}(\omega)\equiv G^{R}(\omega)-G^{R}(\omega)|_{T=0}$. 
Due to this subtraction, the ultraviolet behavior of $G^R$ is improved so that the contribution from the arc with infinite radius becomes negligible.
Another subtraction of $\delta G^R_\infty\equiv \delta G^R(i\omega)|_{\omega\rightarrow \infty}$ is for removing any possibly remaining ultraviolet divergence. 
Taking the infinitesimal $\omega$ limit, we get 
\begin{align}
\label{eq:residue-omega0}
\delta G^{R}(0)-\delta G^{R}_\infty
&= \frac{2}{\pi}\int^\infty_{0} \frac{d\omega}{\omega} \delta \rho(\omega),
\end{align}
where we have made use of the fact that 
the real (imaginary) part of $G^R(\omega)$ is an even (odd) function of $\omega$, and introduced the spectral 
function, $\rho(\omega)\equiv {\text{Im}}G^R(\omega)$.
We also changed the integration variable to $\omega$ for simplicity. 

On the left-hand side, the ultraviolet (UV) and infrared (IR) limits of $G^R$ constrain the spectral function integral through Eq.~(\ref{eq:residue-omega0}). 
The former quantity can be evaluated using the operator product expansion (OPE)~\cite{CaronHuot:2009ns, Shifman:1978by}. 
Because of the subtraction of the $T=0$ piece, all terms with operators of mass dimensions less than four vanish, 
so that the asymptotic behavior at large $\omega$ is described by the operators with mass dimensions four. 
By computing the coefficients of such operators at leading order in $\alpha_s$, we get 
\begin{align} 
\label{eq:GR-OPE-free}  
\begin{split}
\delta G^R(\omega)&= e^2\sum_f q^2_f\frac{1}{\omega^2}
\Biggl[2m_f\delta \langle \overline{\psi}_f \psi_f\rangle
+\frac{1}{12}\delta\left\langle \frac{\alpha_s}{\pi}G^2 \right\rangle \\
&~~~+\frac{8}{3}\delta\langle T^{00}_{f}\rangle\Biggr] 
+{\cal O}\left(\omega^{-4}\right),
\end{split} 
\end{align} 
where $G^{\mu\nu}_a\equiv \partial^\mu A^\nu_a-\partial^\nu A^\mu_a -gf_{abc} A^\mu_b A^\nu_c$ is the field strength, 
$G^2\equiv G^a_{\mu\nu}G^{a\mu\nu}$, $T^{\alpha\beta}_{f}\equiv i{\cal {ST}} \overline{\psi}_f\gamma^\alpha D^\beta \psi_f $ is 
the quark component to the traceless part of the energy-momentum tensor, $D^\mu\equiv \partial^\mu +igA^\mu_a t^a$ the covariant derivative, 
$A^\mu_a$ the gluon field, $t^a$ the generator of the $SU(\Nc)$ group in the fundamental representation, $f_{abc}$ the 
structure constant of the $SU(\Nc)$ group, $m_f$ the current quark mass, $g$ the QCD coupling constant, $\alpha_s\equiv g^2/(4\pi)$, and $\Nc$ the number of the colors. 
${\cal {ST}}$ makes a tensor symmetric and traceless: 
${\cal {ST}}O^{\alpha\beta}\equiv (O^{\alpha\beta}+O^{\beta\alpha})/2-g^{\alpha\beta}O^\mu_\mu /4$. 
We note that having dropped higher order corrections to the coefficients above will be justified in the $\omega\rightarrow \infty$ limit, which allows us to use asymptotic freedom. 
Also note that the traceless gluonic component of the energy-momentum tensor 
[$T^{00}_g$, defined above Eq.\,(\ref{eq:scaling-T'-Ttilde})] can also in principle appear in the OPE at finite temperature. 
We have dropped such a term since it vanishes at leading order in $\alpha_s$, but we will discuss below that it shows up once the operator mixing is taken into account.
We retained the gluon condensate term though formally it is of higher order in $\alpha_s$, as it turns out to be finite even in the $\omega\rightarrow \infty$ limit 
 due to its vanishing anomalous dimension. 
When considering the $\omega\rightarrow \infty$ limit, we need to take into account the effects of scaling and mixing of the operators, reflected in their anomalous dimensions.
The anomalous dimensions of the chiral and gluon condensates are zero, so they do neither scale nor mix. 
On the other hand, the quark energy momentum tensor both scales and mixes with a respective gluonic operator. 
To understand this behavior, we rewrite the operator as $T^{00}_{f}=T'{}^{00}_{f}+(T^{00}+2\tilde{T}^{00}/N_f)/(4C_F+N_f)$, 
where $T'{}^{00}_{f}\equiv T^{00}_{f}-\sum_{f'} T^{00}_{f'}/N_f$, $T^{00}\equiv \sum_{f'} T^{00}_{f'}+T^{00}_g$, 
and $\tilde{T}^{00}\equiv 2C_F  \sum_{f'} T^{00}_{f'}-N_f T^{00}_g /2$. 
Here, $T^{\mu\nu}_g\equiv -G^{\mu\alpha}_{a}G^\nu{}_{\alpha a} +g^{\mu\nu}G^2/4$ is the gluon component of the traceless part of the energy-momentum tensor, 
$N_f$ the flavor number, and $C_F\equiv (\Nc^2-1)/(2\Nc)$. 
A standard renormalization group (RG) analysis yields the following scaling properties~\cite{Peskin:1995ev}: 
\begin{align}
\label{eq:scaling-T'-Ttilde}
\begin{split} 
T'{}^{00}_{f}(\kappa)&=  \left[\frac{\ln\left(\kappa_0/\Lambda_{\text{QCD}} \right)}{\ln\left(\kappa/\Lambda_{\text{QCD}} \right)}\right]^{a'} 
T'{}^{00}_{f}(\kappa_0),\\
 \tilde{T}^{00}(\kappa)
&= \left[\frac{\ln\left(\kappa_0/\Lambda_{\text{QCD}} \right)}{\ln\left(\kappa/\Lambda_{\text{QCD}} \right)}\right]^{\tilde{a}}
 \tilde{T}^{00}(\kappa_0),
\end{split}
\end{align}
while $T^{00}$ is independent of $\kappa$. 
Here $\kappa$ and $\kappa_0$ are renormalization scales, $\Lambda_{\text{QCD}}$ is the QCD scale parameter, 
$a'\equiv 8\Cf/(3b_0)$, and $\tilde{a}\equiv 2(4\Cf+\Nf)/(3b_0)$, where $b_0\equiv (11\Nc-2\Nf)/3$, 
which appears in the expression $\alpha_s(\kappa)=2\pi/[b_0(\ln(\kappa/\Lambda_{\text{QCD}}))]$.
We see that, except for the $T^{00}$ term, 
all terms are suppressed logarithmically at large $\omega$. 
Thus, the resultant expression becomes
\begin{align} 
\label{eq:GR-OPE}
\begin{split}
\delta G^R(\omega)&= e^2\sum_f q^2_f\frac{1}{\omega^2}
\Biggl[2m_f\delta \langle \overline{\psi}_f \psi_f\rangle
+\frac{1}{12}\delta\left\langle \frac{\alpha_s}{\pi}G^2 \right\rangle \\
&~~~+\frac{8}{3}\frac{\delta\langle T^{00}\rangle}{4C_F+N_f}\Biggr] .
\end{split} 
\end{align} 
This vanishes at $\omega \to \infty$ and hence its contribution to Eq.~(\ref{eq:residue-omega0}) is zero.
We note that, in $\omega\rightarrow \infty$ limit, which is relevant to the derivation of the sum rule, the asymptotic freedom of QCD guarantees 
that the above expression is exact. 

On the other hand, the IR limit is well described by hydrodynamics.
At $|\vp|=0$, it suffices to consider the constitutive relation for the system at rest, $\vj= \sigma\vE -\sigma\tau_J \partial_t\vE+{\cal O}(\partial^{2}E)$, 
since the conservation law of the current is trivial ($\partial_t j^0=-\nabla\cdot\vj=0$).
Here $\sigma$ is the electrical conductivity, $\tau_J$ the second order transport coefficient for $\partial_t\vE$, $\vE\equiv -\nabla A^0-\partial_0 \vA$ the electric field, and $A^\mu$ the vector potential. 
We have dropped magnetic field dependent terms and the diffusion term from the constitutive relation, since they vanish in the $|\vp|=0$ case. 
The linear response theory enables us to extract the retarded function through the relation, 
\begin{align}
\label{eq:linear-response} 
j_\mu(\omega)= -G^{R}_{\mu\nu}(\omega) A^\nu(\omega),
\end{align}
which results in
\begin{align}
\label{eq:GR-hydro}
\begin{split}
G^R(\omega)
=i\omega\sigma\left(1+i\tau_J \omega \right)
+{\cal O}(\omega^3 ),~~
\rho(\omega)
&=\sigma\omega
+{\cal O}(\omega^3 ). 
\end{split}
\end{align}
To get $\delta G^R$, we need to know $G^R|_{T=0}$.
 Lorentz invariance guarantees the following form: 
\begin{align}
G^R(\omega)|_{T=0}
&= \omega^2 G_2(\omega^2). 
\end{align}
Here the real part of $G_2$ contains a UV divergence coming from the $T=0$ part, so the renormalization of the photon wave function~\cite{Peskin:1995ev, Bali:2014kia} is necessary, which implies $G_2(0)=0$.
We note that $\sigma$ and $\tau_J$ in Eq.~(\ref{eq:GR-hydro}) need to be defined for the renormalized version of $G^R(\omega)$.
Also the imaginary part of $G_2$ at small $\omega$ is zero because even the lightest vector meson (the $\rho$ meson) has non-zero mass and its spectral strength vanishes below the $\pi \pi$ threshold, so that the spectral weight around $\omega=0$ is zero.
For these two reasons, we see that the left-hand side of Eq.~(\ref{eq:GR-hydro}) is actually equal to $\delta G^R$.
This is not the case for the higher order terms that are of order $\omega^4$ or higher.

We also note that Eq.~(\ref{eq:GR-hydro}) is correct only in the large $\Nc$ limit, in which the coupling effect among the hydro modes is negligible~\cite{Kovtun:2003vj}.
Beyond this limit, a nonanalytic term ({$\sim$}$\omega^{3/2}$) appears in $\rho(\omega)$.
Nevertheless, this does not affect the sum rules 1 and 2 we derive in this work. 
Applying the UV and IR results of Eqs.~(\ref{eq:GR-OPE}), (\ref{eq:GR-hydro}), Eq.~(\ref{eq:residue-omega0}) becomes 
\begin{align}
\label{eq:sumrule-1}
0&= \int^\infty_{0} \frac{d\omega}{\omega} \delta \rho(\omega).
\end{align}
This is the first sum rule (sum rule 1) to be discussed in this paper. 
We should mention here that this is the $|\vp|=0$ version of the sum rule derived in Ref.~\cite{Bernecker:2011gh}. 

Let us check that this sum rule is satisfied at weak coupling and in the chiral limit ($m_f=0$). 
In this case, the spectral function consists of a transport peak at low energy ($\omega\sim g^4T$) and a continuum in the high energy region ($\omega\sim T$). 
We first evaluate the former contribution, which 
can be described by the Boltzmann equation 
\begin{align}
\begin{split} 
&Dn_{\pm f}(\vk, X) 
-\tau^{-1} \nf(|\vk|) \\
&~~~= \mp eq_f\left(\vE+\vv\times\vB\right)(X) \cdot\nabla_{\vk} n_{\pm f}(\vk, X), 
\end{split}
\end{align}
where $D\equiv v\cdot\partial_X+\tau^{-1} $, $n_{\pm f}(\vk, X)$ is the distribution function for the quark (anti-quark) with 
momentum $\vk$ at point $X$, $\nf(|\vk|)\equiv [\exp (|\vk|/T)+1]^{-1}$ is the distribution function at equilibrium, and $v^\mu\equiv (1,\vv)$ with $\vv\equiv \vk/|\vk|$. 
We have adopted here the relaxation time approximation, which considerably simplifies the collision term\footnote{This is a very simple approximation, which was however found to work well 
by solving the Boltzmann equation without relying on it~\cite{Moore:2006qn}: 
The full calculation produces a solution for $\rho(\omega)/\omega$ that is an almost perfect Lorentzian, which is also obtained by the relaxation time approximation 
[see Eq.~(\ref{eq:spectrum-Boltzmann})].}. 
$\tau$ is called relaxation time, and its order of magnitude is determined by collision effects. 
Since we are interested in the retarded Green function, we only need the linearized equation:
$D\delta n_{\pm f}(\vk, X) 
= \mp eq_f \vE(X) \cdot\vv \nf'(|\vk|)$, where $\delta n_{\pm f}\equiv n_{\pm f}-\nf$.
After performing the Fourier transformation $X\rightarrow p$ and setting $|\vp|=0$, this results in the solution
\begin{align}
\label{eq:solution-Boltzmann}
\delta n_{\pm f}(\vk, \omega) 
&= \mp i e q_f \frac{\vE(\omega) \cdot\vv}{\omega+i\tau^{-1}} \nf'(|\vk|).
\end{align}
The induced current is given by
$\vj(\omega)
= 2 e\Nc \sum_f q_f 
\int d^3\vk\vv \sum_{s=\pm 1} s\delta n_{s f}(\vk, \omega)/(2\pi)^3$, where the factor 2 comes from the spin degeneracy of the quarks. 
This expression and Eq.~(\ref{eq:solution-Boltzmann}), together with the linear response relation of Eq.~(\ref{eq:linear-response}), 
give us the following result for $G^R$ and the spectral function:
\begin{align}
\label{eq:spectrum-Boltzmann}
\begin{split}
G^R(\omega)
=& -\frac{T^2 C_{\mathrm{em}}\Nc}{9}
\frac{\omega }{\omega+i\tau^{-1}}, \\
\rho(\omega)
=& \frac{T^2 C_{\mathrm{em}} \Nc}{9}
\frac{\omega \tau^{-1}}{\omega^2+\tau^{-2}}.
\end{split}
\end{align}
Here we have introduced the factor, $C_{\mathrm{em}} \equiv e^2 \sum_f q^2_f$. 
We note that this is reduced to Eq.~(\ref{eq:GR-hydro}) when $\omega\ll \tau^{-1}$, and we can identify
$\sigma= T^2 C_{\mathrm{em}} \Nc \tau/9$ and $\tau_J=\tau$. 
We note that collisional effects are essential 
for the evaluation of $\sigma$: 
If we take the $\tau^{-1}\rightarrow 0$ limit, $\rho(\omega)$ will be proportional to $\omega\delta(\omega)$ and $\sigma$ proportional to $\tau$, which is infinitely large. 
This abnormal behavior indicates that collisions are important in the small $\omega$ region.  
Its contribution to the sum rule Eq.~(\ref{eq:sumrule-1}) reads 
\begin{align} 
\label{eq:contribution-1-hydro}
\int^\infty_{0} \frac{d\omega}{\omega} \delta \rho(\omega)
&= \frac{\pi T^2 C_{\mathrm{em}} \Nc}{18},
\end{align}
which is of order $e^2T^2$, and independent of $\tau$.
Here the $T=0$ component does not contribute because of the absence of the transport peak in the vacuum. 

Next, we evaluate the contribution from the continuum.
From a one-loop calculation~\cite{Altherr:1989jc}, we have 
\begin{align}
\label{eq:spectrum-continuum}
\rho(\omega)
&= \frac{\Nc C_{\mathrm{em}} }{12\pi} \omega^2
\left(1-2\nf\left(\frac{\omega}{2}\right)\right).
\end{align} 
The pair creation/annihilation process of the quark and the anti-quark is responsible for this expression:
one can see that by rewriting the distribution function factor $1-2\nf$ as $[1-\nf]^2-\nf^2$.
The former (latter) term comes from the pair creation (annihilation) process.
It is noted that, after subtracting the $T=0$ part, the spectral function becomes negative. 
Performing the integral over $\omega$, it 
is straightforward to see that its contribution to the sum rule cancels the contribution from the transport peak, Eq.~(\ref{eq:contribution-1-hydro}), 
so that the sum rule Eq.~(\ref{eq:sumrule-1}) is satisfied.

\subsection{\label{SubsectionSumRule2} Sum Rule 2}
In a similar way (replacing $G^R$ with $\omega^2 G^R$ in the derivation), we derive another sum rule which contains two more powers of $\omega$ in the integrand. 
In analogy to the derivation of sum rule 1, we get $\delta G^{R2}_0-\delta G^{R2}_\infty=2 \int^\infty_{0} d\omega\omega \delta \rho(\omega)/\pi$, 
where $\delta G^{R2}_\infty\equiv \omega^2\delta G^R(\omega)|_{\omega\rightarrow \infty}$ and 
$\delta G^{R2}_0\equiv \omega^2\delta G^R(\omega)|_{\omega\rightarrow 0}$.
By using the UV/IR limits of $G^R$, Eqs.~(\ref{eq:GR-OPE}) and (\ref{eq:GR-hydro}), we obtain 
\begin{align} 
\label{eq:sumrule-2} 
\begin{split}
 \frac{2}{\pi} \int^\infty_{0} d\omega\omega & \delta \rho(\omega) = 
 -e^2\sum_f q^2_f
\Bigl[2m_f\delta \langle \overline{\psi}_f \psi_f\rangle \\
& +\frac{1}{12}\delta\left\langle \frac{\alpha_s}{\pi}G^2 \right\rangle 
+\frac{8}{3(4\Cf+\Nf)} \delta\langle T^{00}\rangle\Bigr] .
\end{split}
\end{align} 
This is the second sum rule (sum rule 2)\footnote{We note that this sum rule in the case of $N_f=1$ and $N_c=3$ was derived in Ref.~\cite{Huang:1994fs}.
However, the coefficient of $T^{00}$ in this reference is not the correct one (Eq.~(\ref{eq:sumrule-2})), but is equal to that in the expression (Eq.~(\ref{eq:sumrule-2-free})), where the effect of the mixing/rescaling of the energy-momentum tensor is neglected.}
we discuss in this work. 

\begin{figure}[t] 
\begin{center}
\includegraphics[width=0.45\textwidth]{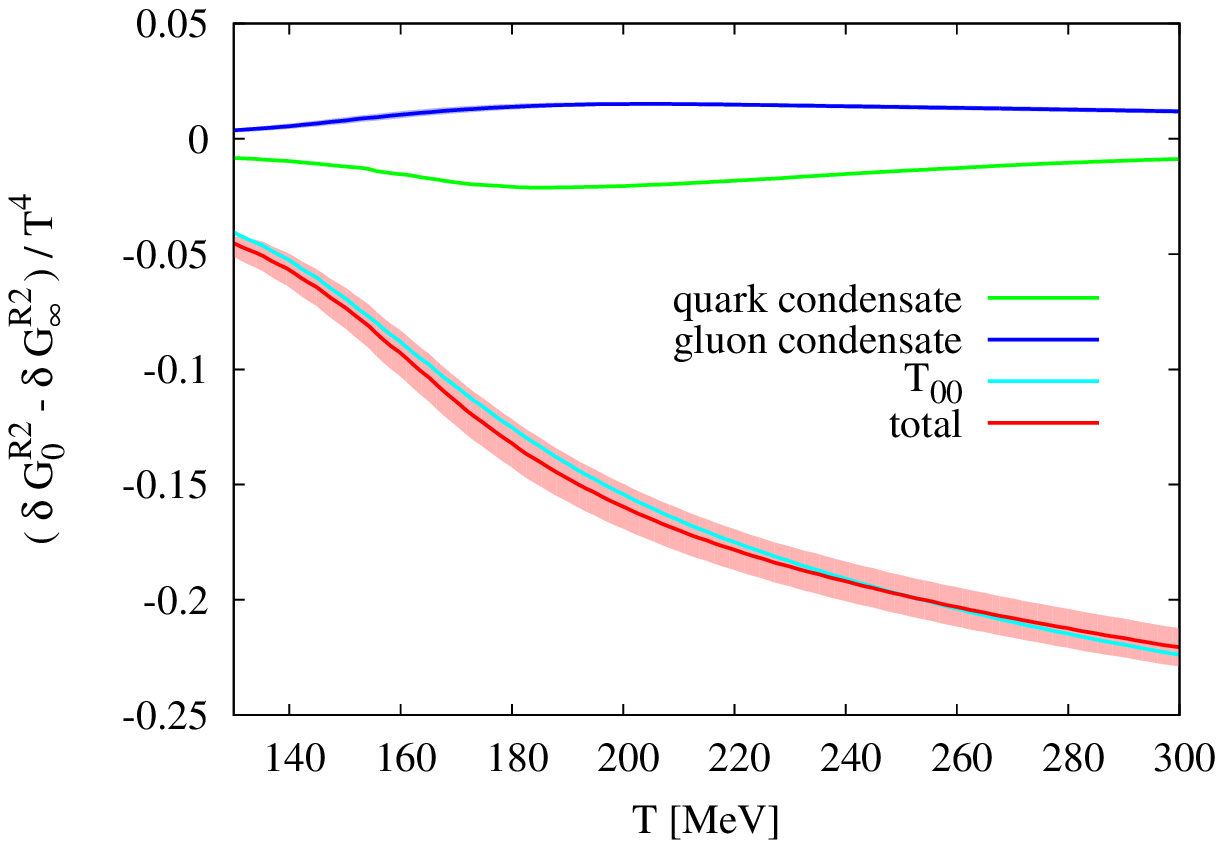}
\includegraphics[width=0.45\textwidth]{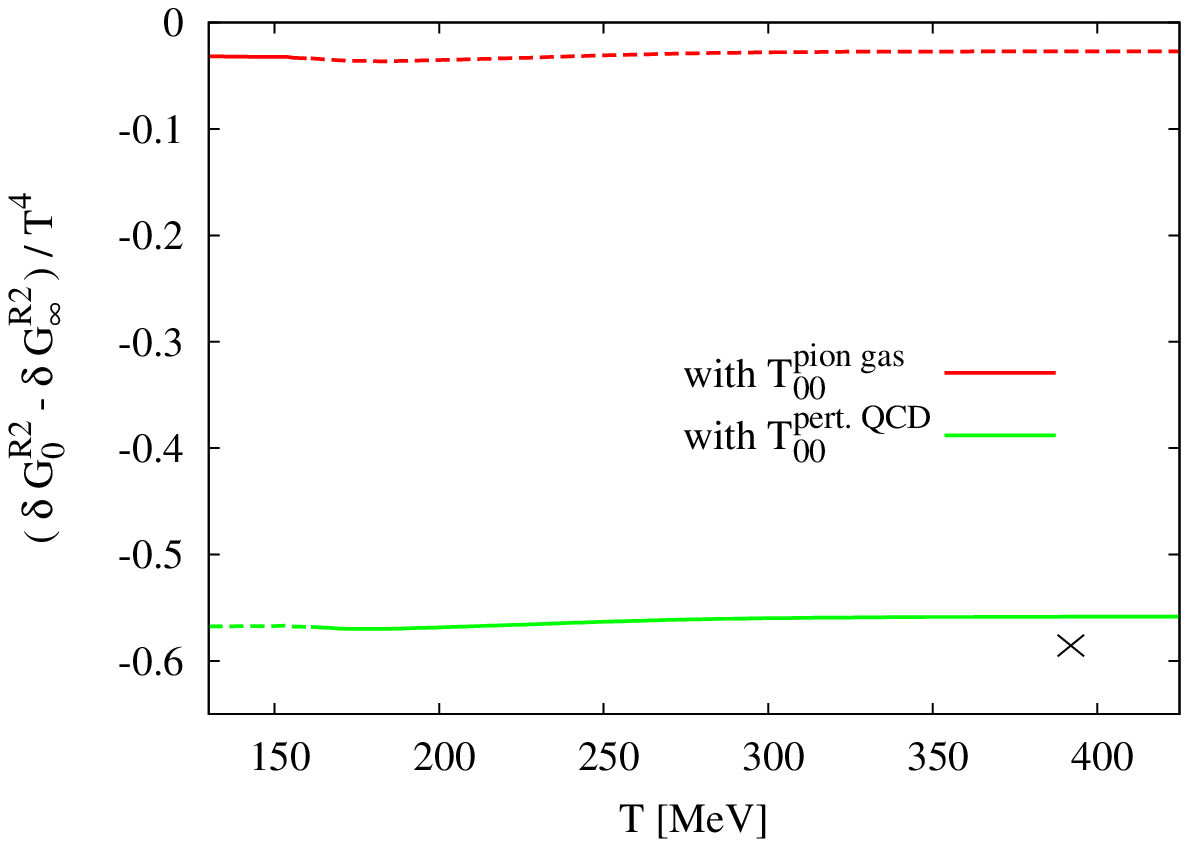}
\vspace{10mm}
\caption{The right-hand sides of Eq.~(\ref{eq:sumrule-2}) (upper plot) and Eq.~(\ref{eq:sumrule-2-free}) (lower plot), divided by $T^4$ and 
shown as a function of temperature $T$. To extract the temperature dependence of the condensates, lattice QCD data provided in 
Ref.~\cite{Bazavov:2014pvz} were used. $T^{00}_f$ in Eq.~(\ref{eq:sumrule-2-free}) was estimated within a free pion gas model, 
reliable at low $T$, and leading order perturbative QCD, which should give the correct behavior at high $T$. 
We used the value $e^2=0.092$ for the plots.
For the free pion gas model, we employ the free pion mass averaged over the three isospin states, $m_\pi=138$MeV. 
The cross at the lower 
right side of the lower plot marks the of our fitted spectral function given in Eq.~(\ref{eq:sumrule2.integral}). For details, see the main 
text of Section \ref{ApplLatticeQCD}.
} 
\label{fig:condensate} 
\end{center} 
\end{figure}

It should be emphasized here that the condensates appearing on the right-hand side of this sum rule are static quantities, 
that can be evaluated non-perturbatively from lattice QCD. 
The gluon condensate can be computed by using the relation at leading order in $\alpha_s$, 
\begin{align} 
e-3p
&=\sum_f m_f\delta \langle \overline{\psi}_f \psi_f\rangle 
-\frac{11\Nc-2\Nf}{24}\delta\left\langle \frac{\alpha_s}{\pi}G^2 \right\rangle. 
\end{align} 
We note that though the sum rule (\ref{eq:sumrule-2}) is exact, the evaluation of the gluon condensate using the expression above is valid only perturbatively. 
In this study, we take the 
chiral condensate, energy and pressure from a recent $\Nf=2+1$ lattice calculation by the HotQCD Collaboration~\cite{Bazavov:2014pvz}. 
To understand the behavior of the different terms on the 
right-hand side of Eq.~(\ref{eq:sumrule-2}), they are shown in the first plot of Fig.~\ref{fig:condensate} as a function of temperature. 
It is seen in this figure that the quark and gluon condensate terms are relatively small, comparable in magnitude and have opposite signs. 
Their contributions therefore cancel to a large degree, so that the right-hand side of Eq.~(\ref{eq:sumrule-2}) is almost completely 
determined by the dominant $\delta \langle T^{00} \rangle$ term. 
Because this term does not depend on quark flavor, the decomposition of Eq.~(\ref{eq:sumrule-2}) into its 
flavor components is determined simply by the quark charges $q_f$, meaning that the $u$-quark contribution is about 
a factor of four larger than those of the $d$ and $s$-quarks. 

For future convenience, we also write the expression obtained by neglecting the scaling and mixing of the operators, which is obtained from Eq.~(\ref{eq:GR-OPE-free}):
\begin{align} 
\label{eq:sumrule-2-free} 
\begin{split}
& \frac{2}{\pi} \int^\infty_{0} d\omega\omega \delta \rho(\omega)= \\
&~~~ -e^2\sum_f q^2_f
\left[2m_f\delta \langle \overline{\psi}_f \psi_f\rangle
+\frac{1}{12}\delta\left\langle \frac{\alpha_s}{\pi}G^2 \right\rangle 
+\frac{8}{3} \delta\langle T^{00}_{f}\rangle\right] .
\end{split}
\end{align}  
Lattice data for $\delta\langle T^{00}_{f}\rangle$ are not available at present, so that we need 
to employ model to estimate the condensates in Eq.~(\ref{eq:sumrule-2-free}). 
This is done by using a free pion/quark gas model, which is reliable at small/large $T$. 
In the former model, $\delta\langle T^{00}_{f}\rangle$ is evaluated as~\cite{Zschocke:2002mn}
$\delta\langle T^{\alpha\beta}_{f}\rangle
= 3 \int d^3\vk \nb(E_k) 
 \left(k^\alpha k^\beta-k^2 g^{\alpha\beta}/4\right)
 A^f_{2,\pi}(\kappa^2)/[(2\pi)^3 2E_k]$, where $E_k\equiv\sqrt{\vk^2+m^2_\pi}$, $m_\pi$ is the pion mass, and $\nb(E)\equiv [e^{E/T}-1]^{-1}$. 
$A^f_{2,\pi}(\kappa^2)$ are moments of the quark distribution functions of quarks and anti-quarks in the pion at scale $\kappa^2$: 
$A^f_{2,\pi}(\kappa^2)\equiv 2\int^1_0 dx x [q^f_\pi(x,\kappa^2)+\overline{q}^f_\pi(x,\kappa^2)]$. 
We will use $A^{u+d}_{2,\pi}(1\,\mathrm{GeV}^2) = 0.97$ \cite{Hatsuda} and ignore the strange quark contribution, which 
is about an order of magnitude smaller than that of the $u$ and $d$ quarks \cite{Asakawa}. 
We then have 
\begin{align}
\delta\langle T^{00}_{f}\rangle 
= \frac{3}{2} A^f_{2,\pi}(\kappa^2)\left[\frac{3}{4}m^2_\pi I^\pi_1+I^\pi_2 \right],
\end{align}
with $I^\pi_n\equiv \int^\infty_0 d|\vk| |\vk|^{2n} \nb(E_k)/(2\pi^2 E_k)$. 
On the other hand, in the latter model, the results reads
\begin{align}
\label{eq:STqq-free}
\delta\langle T^{00}_{f} \rangle
&= \Nc\frac{7\pi^2T^4}{60}.
\end{align}
Both results are plotted in the second plot of Fig.~\ref{fig:condensate}, combined with the other terms 
of the right-hand side of Eq.~(\ref{eq:sumrule-2-free}). 
From this figure, one can see that 
the shown function must have an increasing behavior as the temperature 
is changed from below $T_c$, where pions dominate, to temperatures where perturbative QCD becomes 
reliable. 

We can check that the sum rule 2 is satisfied at weak coupling and in the chiral limit. 
First, it is noted that the contribution from the continuum to the integral of the spectral function dominates over that from the transport peak. 
Using Eq.~(\ref{eq:spectrum-continuum}), the continuum contribution is evaluated as $-14\pi^2T^4 C_{\mathrm{em}} \Nc /45$. 
The contribution from the low energy region is much smaller:
It is estimated by using Eq.~(\ref{eq:spectrum-Boltzmann}) as 
\begin{align}
\begin{split}
  \frac{2}{\pi} \int^\Lambda_{0} d\omega\omega \delta \rho(\omega)
&= \frac{2}{\pi}\frac{T^2 C_{\mathrm{em}}}{9} \tau^{-1}\Nc\int^\Lambda_{0} d\omega
\frac{\omega^2 }{\omega^2+\tau^{-2}} \\
&\sim e^2T^2\tau^{-1} \Lambda ,
\end{split}
\end{align}
where $\Lambda$ is the UV cutoff of the transport peak. 
The Boltzmann equation is applicable when $\omega \ll gT$ since the instantaneous scattering description breaks down~\cite{Moore:2006qn}, which gives the UV cutoff $\Lambda\sim gT$. 
By using this order estimate, we find that the transport peak contribution is much smaller than the continuum contribution $\sim e^2 T^4 $ because of $\tau^{-1}\sim g^4T$.
On the other hand, the condition $m_f=0$ eliminates the chiral condensate term in the sum rule. 
At leading order in $g$, the gluon condensate also vanishes and the right-hand side of Eq.~(\ref{eq:sumrule-2-free}) is reduced to $ -14 \pi^2 T^4 C_{\mathrm{em}} \Nc /45$ by 
using Eq.~(\ref{eq:STqq-free}). 
From these two expressions, one sees that the contribution from the continuum 
leads to a satisfied Eq.~(\ref{eq:sumrule-2-free}), which is the sum rule without operator scaling or mixing, not 
the correct one of Eq.~(\ref{eq:sumrule-2}). 
In fact, it was shown that a two-loop contribution yields an additional structure in the spectral function, 
namely a power-suppressed tail in the UV region ($\omega\gg T$)~\cite{CaronHuot:2009ns}:
\begin{align}
\label{eq:UV-tail}
\begin{split}
\delta\rho(\omega)
\simeq e^2\sum_f q^2_f\frac{1}{\omega^2}
\frac{8}{9}\alpha_s(\omega) & \Bigl[2\Cf \delta\langle T'{}^{00}_{f}(\omega)\rangle \\ 
& +\frac{1}{\Nf}\delta\langle \tilde{T}{}^{00}(\omega)\rangle \Bigr].
\end{split}
\end{align}
By considering the operator scaling effect of Eq.~(\ref{eq:scaling-T'-Ttilde}), and the running coupling, 
the contribution of this UV-tail to the sum rule is found to be 
\begin{align}
\label{eq:contribution-UVtail}
\begin{split} 
&  \frac{2}{\pi} \int^\infty_{\omega_{\text{min}}} d\omega\omega \delta \rho(\omega) \\
&~~~=e^2\sum_f q^2_f \frac{32}{9 b_0} \\ 
&~~~~~~ \times \int^\infty_{X_{\text{min}} } dX
\frac{1}{\Nf}\delta\langle \tilde{T}{}^{00}(\kappa_0)\rangle 
\left[\ln\left(\kappa_0/\Lambda_{\text{QCD}} \right)\right]^{\tilde{a}} \frac{1}{X^{\tilde{a}+1}} \\
&~~~=e^2\sum_f q^2_f \frac{8}{3 }  
\frac{4\Cf \delta\langle T^{00}_f(\omega_{\text{min}})\rangle-\delta\langle T^{00}_g(\omega_{\text{min}})\rangle }{4\Cf+\Nf},
\end{split}
\end{align}
where $X\equiv \ln(\omega/\Lambda_{\text{QCD}})$, $X_{\text{min}}\equiv \ln[\omega_{\text{min}}/\Lambda_{\text{QCD}}]$, 
and $\omega_{\text{min}}\sim T$ is the IR cutoff of the tail. 
We have furthermore made use of the fact that 
that $\langle T{}^{00}_{f}\rangle$ does not depend on $f$ in the chiral limit. 
Applying the expression of the energy density in the free and massless limit [Eq.~(\ref{eq:STqq-free})], the sum of the contributions from the UV tail of 
Eq.~(\ref{eq:contribution-UVtail}) and the continuum ($-e^2\sum_f q^2_f 8 \delta\langle T^{00}_f\rangle /3 $) 
is found to be $-C_{\mathrm{em}} 8 \delta\langle T^{00} \rangle /[3(4\Cf+\Nf)]$.
This is nothing but the right-hand side of Eq.~(\ref{eq:sumrule-2}), which demonstrates that the sum rule is satisfied only considering the contribution of the UV tail in the above limits. 
Also, it is easy to see that such contribution to the other two sum rules is negligible at weak coupling (with order estimate $\omega_{\text{min}}\sim T$).

\subsection{Sum Rule 3}
In the sum rule to be discussed in this subsection, the integrand of sum rule 1 is in essence divided by $\omega^2$. 
To avoid potential IR divergences, the derivation however has to be carried out with some care. 
Equation~(\ref{eq:residue-theorem}) can be written as
\begin{align}
\begin{split}
\delta G^{R}(i\omega)-\delta G^{R}_\infty
&= \frac{1}{2\pi }\int^\infty_{-\infty} d\omega' 
\frac{1}{\omega'{}^2+\omega^2} \\
&~~~\times(\omega' \delta \rho(\omega')+\omega {\text{Re}}[\delta G^{R}(\omega')-\delta G^{R}_\infty])\\
&=\frac{1}{\pi }\int^\infty_{-\infty} d\omega' 
\frac{\omega' \delta \rho(\omega')}
{\omega'{}^2+\omega^2},
\end{split}
\end{align}
where in the second line we have used the property that the contributions from the first and the second terms are equal, which can be shown 
by evaluating the right-hand side of Eq.~(\ref{eq:residue-theorem}) using the residue theorem with the contour closing in the lower half plane. 
Subtracting Eq.~(\ref{eq:residue-omega0}) and $-\sigma\omega$ from this expression and using Eq.~(\ref{eq:GR-hydro}) on the left-hand side, we get 
\begin{align}
\label{eq:sumrule-3}
-\sigma\tau_J
&= \frac{2}{\pi} \int^\infty_{0} \frac{d\omega}{\omega^3} 
\left[\delta \rho(\omega)-\sigma\omega\right],
\end{align}
in which the $-\sigma\omega$ term in the integrand is included to remove the IR 
singularity. 
This is the third sum rule (sum rule 3) we have derived in this paper.

Let us again check that this sum rule is satisfied at weak coupling. 
The contribution from the transport peak is found to be
\begin{align}
\label{eq:tranport.peak.3}
\frac{2}{\pi} \int^\infty_{0} \frac{d\omega}{\omega^3} 
\left[\delta \rho(\omega)-\sigma\omega\right]
&=-\frac{T^2 C_{\mathrm{em}}}{9}\Nc\tau^2,
\end{align} 
where we have used 
Eq.~(\ref{eq:spectrum-Boltzmann}) and the expression of $\sigma$ in the relaxation time approximation.
Taking into account $\tau_J = \tau$, 
we find that Eq.~(\ref{eq:tranport.peak.3}) is equal to the left-hand side of sum rule 3. 
The contribution from the continuum is much smaller than that from the transport peak, due to the negative power of $\omega$ in the integrand:
From Eq.~(\ref{eq:spectrum-continuum}), the continuum contributes to the sum rule as 
$2 \int^\infty_{\mu} d\omega \left[\delta \rho(\omega)-\sigma\omega\right] /(\pi\omega^3) \sim \sigma/\mu$, where $\mu$ is the IR cutoff.
The one-loop result of Eq.~(\ref{eq:spectrum-continuum}) is reliable for $\omega \gg gT$, while for $\omega \leq gT$ the effect of thermal modification 
of the quark spectrum and the vertex becomes important so that the hard thermal loop resummation is necessary~\cite{Moore:2006qn, Braaten:1990wp}. 
It is thus natural to set the IR cutoff to $\mu\sim gT$. 
With this order estimate, the contribution from the continuum turns out to be much smaller than $-\sigma\tau_J\sim e^2 g^{-8}$. 

A comment on the sensitivity on the continuum/transport peak of the sum rules is in order here. 
From the discussions above, sum rule 1 was found to be equally sensitive to both of them, at least in the weak coupling regime. 
Meanwhile, sum rule 2 (3) is more sensitive to the continuum (transport peak) because of positive (negative) power of $\omega$ in the integrand. 
This suggests that, if one wishes to extract information of one of these objects from the sum rules, one should use 
the most suitable one, which is most sensitive to the object of interest. 

\section{\label{ApplLatticeQCD} Application to Lattice QCD 
data analysis}

Let us demonstrate that the sum rules we have derived can be used to give constraints to the 
spectral ansatz used in fits to lattice QCD data. 
As a first trial, we consider 
the simple\footnote{A more complicated ansatz, which also contains information on vacuum bound states, was introduced in Refs.~\cite{Brandt,Brandt:2012jc}. 
In these works, the sum rule of Eq.~(\ref{eq:sumrule-1}) was furthermore used to constrain the parameters appearing in their ansatz.}
 ansatz introduced in Ref.~\cite{Ding:2010ga} ({all quantities proportional to 
$\rho(\omega)$ or $G^R$ in this work are multiplied by a factor of $1/6$ compared to the corresponding expressions in Ref.~\cite{Ding:2010ga}}),
\begin{align} 
\label{eq:ansatz-1}
\begin{split}
\rho(\omega)
&= C_{\mathrm{em}}
\left[ c_{BW}\rho_{\text{peak}}(\omega)
+(1+k)\rho_{\text{cont}}(\omega)\right],
\end{split}
\end{align}
where  
\begin{align}
\rho_{\text{peak}}(\omega) &\equiv \frac{1}{3} \frac{\omega\Gamma/2}{\omega^2+(\Gamma/2)^2} , \label{eq:Lorentzian} \\
\rho_{\text{cont}}(\omega) &\equiv \frac{\omega^2}{4\pi }\left(1-2\nf\left(\frac{\omega}{2}\right)\right), 
\label{eq:cont.ansatz-1}
\end{align}
correspond to the transport peak and the continuum in the weak coupling limit. 
We note that, $\delta\rho(\omega)$ can be obtained by subtracting $\rho_{T=0}(\omega)$. 
Data for this function can be obtained from the experimental $(e^+ e^- \to \text{hadrons})$ cross 
section (see for instance the compilation of data given in the particle data group~\cite{Olive}), 
or from zero temperature lattice calculations. 
In this paper, we will however for simplicity confine ourselves to the averaged form $C_{\text{em}} \omega^2(1+k)/(4\pi)$. 
Equation (\ref{eq:ansatz-1}) contains three parameters ($c_{BW}, \Gamma, k$) that need to be determined by fitting the data.
Sum rule 1 of Eq.~(\ref{eq:sumrule-1}) provides a constraint on these parameters: 
\begin{align}
\label{eq:constraint}
c_{BW}=
 (1+k)T^2.
\end{align}
This constraint may be used to reduce the number of fitting parameters in the ansatz. 
Here, we simply check whether 
the values of the parameters obtained from the fit~\cite{Ding:2010ga} satisfy the sum rule. 
The fitted values at $T=1.45T_c$ are $k\simeq 0.047, \Gamma\simeq 2.2T, c_{BW}\simeq 1.2T^2$, which give $1.2T^2$ on the left-hand side of Eq.~(\ref{eq:constraint}) while 
$1.0T^2$ is obtained on the right-hand side.
We see that, even though the agreement is not perfect, the fit satisfies the constraint with reasonable precision. 

Nevertheless, the ansatz Eq.~(\ref{eq:ansatz-1}) can not be applied to the other two sum rules, Eqs.~(\ref{eq:sumrule-2}) and (\ref{eq:sumrule-3}), because 
it would cause a UV divergence in sum rule 2 and an IR divergences in sum rule 3. 
Therefore, to construct a spectral function that can satisfy all three sum rules, 
an improved parametrization is necessary. 
We hence propose the following ansatz (ansatz A): 
\begin{align}
\label{eq:ansatz-A}
\begin{split}
\rho(\omega)
&= C_{\mathrm{em}}
\Bigl[c_{BW}\rho_{\text{peak}}(\omega)[1-A(\omega)] \\ 
&~~~+A(\omega)(1+k)\rho_{\text{cont}}(\omega)\Bigr], 
\end{split}
\end{align}
where $A(\omega) \equiv \tanh(\omega^2/\Delta^2)$. 
For consistency, the spectral function at $T=0$ is modified as $\rho_{T=0}(\omega)=C_{\text{em}} A(\omega)\omega^2(1+k)/(4\pi)$.
As one can easily check, the cutoff function $A(\omega)$ removes all IR and 
UV divergences in Eqs.~(\ref{eq:sumrule-2}) and (\ref{eq:sumrule-3}). 

\begin{figure}[t] 
\begin{center}
\includegraphics[width=0.5\textwidth]{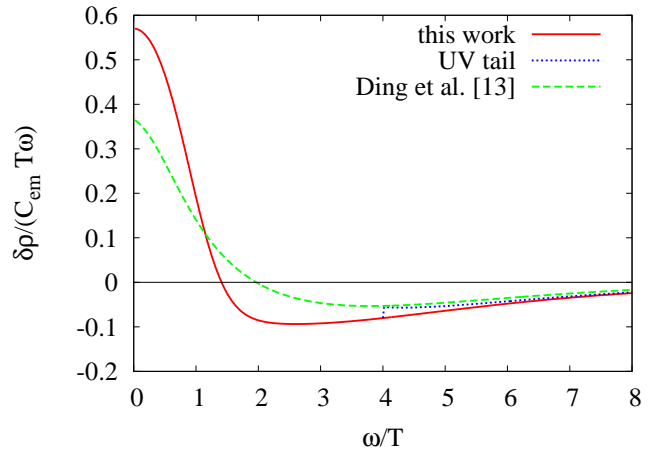}
\vspace{10mm}
\caption{Ansatz A (red solid line), ansatz B (blue dotted line), and the ansatz used in Ref.~\cite{Ding:2010ga} (green dashed line) 
as functions of $\omega$. 
Note that ans$\mathrm{\ddot{a}}$tze A and B are identical for $\omega < \omega_{\mathrm{min}} = 4.0\,T$, where they overlap.
The unit of the vertical axis is $C_{em} T \omega$ while that of the horizontal axis is $T$.
} 
\label{fig:spectral} 
\end{center} 
\end{figure}

\begin{figure}[t] 
\begin{center}
\includegraphics[width=0.5\textwidth]{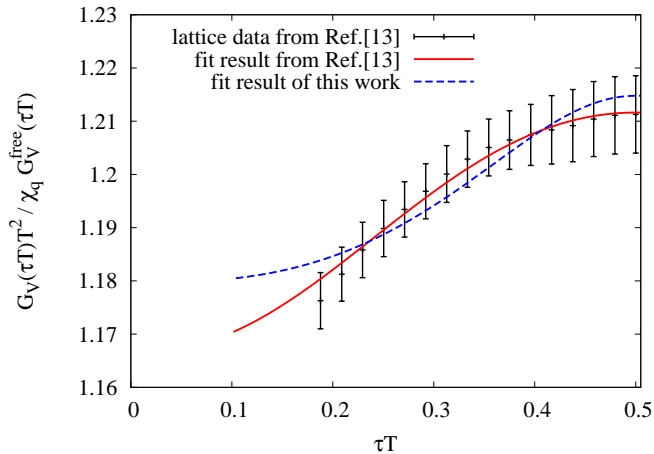}
\vspace{10mm}
\caption{{Lattice data for the Euclidean vector correlator, adapted from Ref.~\cite{Ding:2010ga} (black points), 
the fit result using Eqs.~(\ref{eq:ansatz-1}-\ref{eq:cont.ansatz-1}) (red solid line) and the fit result using the improved functional 
form of Eq.~(\ref{eq:ansatz-A}) (blue dashed line).}} 
\label{fig:GV} 
\end{center} 
\end{figure}

The values of $\Gamma$, $c_{BW}$, $k$, and $\Delta$ should be determined from lattice data. 
To demonstrate that this functional form is feasible, we have performed a simple trial analysis,  
making use of the Euclidean vector correlator and second thermal moment data provided in Ref.\,\cite{Ding:2010ga} 
{for $T = 1.45\,T_c$}. 
These data were also used to fit the ansatz of Eq.~(\ref{eq:ansatz-1}), as explained above. 
{
The Euclidean vector correlator is given in terms of the spectral function as 
\begin{equation}
G^{\mathrm{E}}(\tau, T) = \int_{0}^{\infty} \frac{d \omega}{2 \pi} \rho(\omega) \frac{\cosh[\omega(\tau - 1/2T)]}{\sinh(\omega/2T)},
\label{eq:Eucl.vec.correlator}
\end{equation} 
with Euclidian time $\tau$, while the second thermal moment is defined as
\begin{equation}
G^{(2)}(T) = \frac{1}{2} \int_{0}^{\infty} \frac{d \omega}{2 \pi} \Bigl( \frac{\omega}{T} \Bigr)^2 \frac{\rho(\omega)}{\sinh(\omega/2T)}. 
\label{eq:thermal.moment.1}
\end{equation} 
In Ref.\,\cite{Ding:2010ga} the latter quantity was given relative to its free counterpart: 
\begin{equation}
\frac{G^{(2)}(T)}{G^{(2)}_{\mathrm{free}}(T)} = 1.067 \pm 0.012 \hspace{0.5cm} (T = 1.45\,T_c). 
\label{eq:thermal.moment.2}
\end{equation}
Here, the free second thermal moment can be computed analytically and is in our conventions given as 
\begin{equation}
G^{(2)}_{\mathrm{free}}(T) = \frac{14 \pi^2}{15} T^3. 
\label{eq:thermal.moment.3}
\end{equation}
}
We moreover employ the sum rule 1 of Eq.~(\ref{eq:sumrule-1}) to constrain our fit, as it 
was done in Ref.\,\cite{Brandt,Brandt:2012jc}. 
Specifically, the constraints of the second thermal moment {[Eqs.\,(\ref{eq:thermal.moment.1}) and (\ref{eq:thermal.moment.2})]} and sum rule 1 
{allow us to determine $c_{BW}$ and $k$ and therefore 
to} reduce the number 
of undetermined parameters to two ($\Gamma$ and $\Delta$), which are then fitted to the Euclidean vector correlator data. 
{In this fit, we do not only use the central value of Eq.\,(\ref{eq:thermal.moment.2}), but 
probe the whole range to look for the value that gives the smallest overall $\chi^2$.} 
Following this procedure, we have found that the best fit is obtained for very large values of $\Gamma$, with values of $\Delta/T$ 
of the order of one. This means that the transport peak at low energy is not generated by the Lorentzian of Eq.~(\ref{eq:Lorentzian}), 
but by the function $1 - A(\omega)$, with the width $2\Delta$. As will be shown in Fig.~\ref{fig:spectral}, these two 
functional forms are quite alike and share many qualitative features. 
Quantitatively, our best fit is obtained for, 
\begin{equation}
\begin{split}
k & = 0.058, \\
2 c_{BW} / (T \Gamma) & = 1.7, \\
\Gamma / T & = \text{infinity}, \\
\Delta / T & = 1.2, 
\label{eq:parameters}
\end{split}
\end{equation}
which gives a $\chi^2/d.o.f$ of $0.53$. The respective (vacuum subtracted) spectral function is shown in 
Fig.~\ref{fig:spectral}, together with the fit result of Ref.~\cite{Ding:2010ga}, for which Eq.~(\ref{eq:ansatz-1}) was 
used. 
{To give the reader a better idea on the quality of the fit, we show in Fig.~\ref{fig:GV} 
the Euclidean vector correlator lattice data with our fitted curve. For comparison, we also plot the curve corresponding to 
the fit performed in Ref.~\cite{Ding:2010ga} with Eqs.~(\ref{eq:ansatz-1}-\ref{eq:cont.ansatz-1}). 
$G_{\mathrm{V}}(\tau, T)$, which is used in Fig.~\ref{fig:GV} is defined as $G_{\mathrm{V}}(\tau, T) = G^{\mathrm{E}}(\tau, T) - \chi_q T/6$. 
For the quark number susceptibility $\chi_q$ we employ the value provided in Ref.~\cite{Ding:2010ga}: $\chi_q/T^2 = 0.897$. 
Furthermore, $G^{\mathrm{free}}_{\mathrm{V}}(\tau, T)$ is related to the free Euclidean vector correlator and 
can be given analytically as 
\begin{equation}
\begin{split}
G^{\mathrm{free}}_{\mathrm{V}}(\tau, T) =  T^3 \Biggl[ & \pi(1-2 \tau T) \frac{1 + \cos^2(2 \pi \tau T)}{\sin^3(2 \pi \tau T)} \nonumber \\ 
& + 2 \frac{\cos(2 \pi \tau T)}{\sin^2(2 \pi \tau T)} \Biggr].
\end{split}
\end{equation}
It is seen in Fig.~\ref{fig:GV} that the fit of Ref.~\cite{Ding:2010ga} generally agrees better with the central values of the 
lattice data points. Their errors are however too large to discriminate the two fits. 
Reduced errors and more data points at smaller $\tau T$ values will likely improve this situation and impose 
tougher constraints on the various functional forms used to parametrized the spectral function.}

Having the fitted and well behaved spectral function of Eq.~(\ref{eq:ansatz-A}) at hand, we can now proceed to compute various 
quantities of interest. First of all, one can easily extract the electrical conductivity as 
\begin{equation}
\begin{split}
\frac{\sigma}{T} & = \lim_{\omega \to 0} \frac{\rho(\omega)}{\omega T} = C_{\mathrm{em}} \frac{2 c_{BW}}{3T \Gamma} \\
& = 0.57 \times C_{\text{em}}, 
\label{eq:sigma.value}
\end{split}
\end{equation}
which is about 50\,\% larger than the value reported in \cite{Ding:2010ga}. 

Next, we can check to what degree our spectral function satisfies sum rule 2 of Eq.~(\ref{eq:sumrule-2-free}). 
As we have explained in Section \ref{SubsectionSumRule2}, the difference between the sum rules of Eqs.~(\ref{eq:sumrule-2}) and (\ref{eq:sumrule-2-free}) 
corresponds to a non-exponentially suppressed UV tail of the spectral function. As this tail is not included in the parametrization of Eq.~(\ref{eq:ansatz-A}), its 
integral should be compared to Eq.~(\ref{eq:sumrule-2-free}) and not Eq.~(\ref{eq:sumrule-2}). Computing the integral with the values of Eq.~(\ref{eq:parameters}), 
we get 
\begin{equation}
\frac{2}{\pi} \int^\infty_{0} d\omega\omega \delta \rho(\omega) = - 0. 59 T^4,  
\label{eq:sumrule2.integral}
\end{equation}
with $N_f = 3$. 
This value is marked as a cross on the lower right corner in the second plot of Fig.~\ref{fig:condensate}. 
Note that the lattice data of Ref.~\cite{Ding:2010ga} are given at $1.45\,T_c$. For $T_c$, we have used 
$T_c = 270\,\mathrm{MeV}$, suitable for quenched QCD. 
As can be seen in this plot, the integrated value of Eq.~(\ref{eq:sumrule2.integral}) lies very close to the 
leading order perturbative QCD result, showing that sum rule 2 can be satisfied with reasonable precision, if the 
condensates on its right-hand-side approach the perturbative limit quickly enough. 
To study this question in more detail, an explicit lattice calculation of $\delta \langle T^{00}_{f}\rangle$ will 
however be needed. 

In this context, we note that one could try to construct a spectral function that is consistent with the sum rule of  Eq.~(\ref{eq:sumrule-2}). 
For this purpose, one would need to include the above-mentioned UV tail in the ansatz (ansatz B):
\begin{align}
\label{eq:ansatz-B}
\begin{split}
\rho(\omega)
&= C_{\mathrm{em}}
\Bigl[c_{BW}\rho_{\text{peak}}(\omega)[1-A(\omega)] \\
&+A(\omega)(1+k)\rho_{\text{cont}}(\omega) 
+c_{UV} \theta(\omega-\omega_{\text{min}})\rho_{\text{tail}}(\omega)\Bigr], 
\end{split} 
\end{align}
where 
\begin{align}
\rho_{\text{tail}}(\omega)
&\equiv \frac{4\Cf\pi^2T^4}{9\omega^2}\alpha_s(e\Lambda_{\text{QCD}})[\ln(\omega/\Lambda_{\text{QCD}})]^{-\tilde{a}-1}. 
\label{eq:ansatz-B.2}
\end{align}
$e$ in the above expression stands for Euler's number. 
This form has two extra fitting parameters $\omega_{\text{min}}$ and $c_{UV}$, and would in principle allow us to use the exact sum rule of Eq.~(\ref{eq:sumrule-2}). 
Here $c_{UV}=1$ corresponds to the perturbative result at $m_f=0$, Eq.~(\ref{eq:UV-tail}). 
In this work, we will not pursue this possibility any further and only illustrate the potential effect of $\rho_{\text{tail}}(\omega)$ 
by adding it to our fitted spectral function, using $\Nc=\Nf=3$, $T/\Lambda_{\text{QCD}}=1.5$, $c_{UV}=1$, and $\omega_{\text{min}}= 4.0T$. 
The result is shown as a blue dotted line in Fig.~\ref{fig:spectral}. 
As one can see in this figure, the UV tail just modestly modifies the spectral function in the plotted energy region, at least with the parameters used here. 
It should also be noted that the parametrization of Eq.~(\ref{eq:ansatz-B}) does not provide a completely realistic 
description of the spectral function around $\omega \sim \omega_{\mathrm{min}}$, where it contains a discontinuity. 
Our low energy step-function cutoff however provides the most simple description of the onset of the UV tail with the fewest numbers 
of parameters, and does not cause any divergence in our sum rules.
For these reasons, we have adopted this simple cut-off scheme.
It is also likely to be useful for future spectral function fits to lattice QCD data, which take the UV tail into account.

As a last point, we next discuss the application of sum rule 3 given in Eq.~(\ref{eq:sumrule-3}). 
At first, let us clarify the definition of the parameter appearing in the left-hand side, $\tau_J$. 
It is expressed in terms of the retarded Green function as, $\tau_J\equiv -G^R{}''(\omega=0)/(2\sigma)$ as can be seen from Eq.~(\ref{eq:GR-hydro}). $\tau_J$ therefore 
does not explicitly appear in the spectral function since it corresponds to the real part of $G^R$. 
As the transport coefficient $\tau_J$ is furthermore at present not known, this sum rule can not be used as an additional fitting constraint. 
If the spectral function is however already determined from other sources, Eq.~(\ref{eq:sumrule-3}) can be used to estimate $\tau_J$. 
Using Eqs.~(\ref{eq:ansatz-A}-\ref{eq:sigma.value}), we get 
\begin{equation}
\tau_{J} = 0. 067 C_{\mathrm{em}} / T \hspace{0.5cm} (T = 1.45\,T_c). 
\label{eq:tau.value}
\end{equation}
To our knowledge, this is the first time that this transport coefficient has been determined non-perturbatively. 
Note that the above number is a quenched QCD estimate, 
as we have made use of quenched lattice data to fix the spectral function. 
We do not expect that introducing the UV-tail such as in ansatz B in Eqs.~(\ref{eq:ansatz-B}) and (\ref{eq:ansatz-B.2}) 
will strongly modify the above numerical result for $\tau_{J}$, because the contribution from the high energy part of the spectral function to the sum rule 3 is strongly suppressed, as we have discussed at the end of the previous section.  We have explicitly checked this by computing $\tau_J$ from sum rule 3 of Eq.~(\ref{eq:sumrule-3}), 
using both ans$\mathrm{\ddot{a}}$tze A and B with the parameter 
values given above. As a result, we found that the extracted 
values of $\tau_J$ only differ by about 0.1\,\%, which shows that in practice it 
does not matter which ansatz is used for sum rule 3. 
We should furthermore mention here that, all the ans$\mathrm{\ddot{a}}$tze used in this section do not take into account the 
large $N_c$ suppressed nonanalytic behavior at small $\omega$ (which seems to be challenging to see in current lattice QCD analysis) caused by hydro mode coupling.
It is therefore consistent to use sum rule 3 (\ref{eq:sumrule-3}), which does not consider this effect as well.

\section{Concluding Remarks}

We give a few final comments on future perspectives of this work. 
In this paper we have so far for simplicity only analyzed the zero-momentum ($|\vp|=0$) case. Generalizing our analysis to finite, but small $|\vp|$ is 
straightforward and is worth investigating in detail. 
In this case, one needs to analyze both the longitudinal and the transverse channels separately. 
Apart from that, other transport coefficients such as the diffusion constant and another one related to the magnetic sector will appear in the sum rules. 
Also, since the sum rules are exact, it would be interesting to check their validity by explicit calculations in the hadron phase below $T_c$ and/or the strong 
coupling regime. 
We plan to report on parts of these generalizations in a full publication in the near future. 

\section*{ACKNOWLEDGMENTS}
We thank Wolfram Weise, Guillaume Clement Beuf, Su Houng Lee, and Tetsuo Hatsuda for fruitful discussions and 
in particular thank Wolfram Weise for his careful reading of our manuscript.  
D.S. is supported by the Alexander von Humboldt Foundation.


\end{document}